\documentclass[10pt,
prd,
twocolumn,
nofootinbib,
noshowkeys,
noshowpacs,
superscriptaddress,
floatfix
]{revtex4-2}
\usepackage{amsmath,amsfonts,amsthm,amssymb}
\usepackage[dvips]{graphics,graphicx}
\usepackage[usenames,dvipsnames]{color}
\definecolor{darkblue}{RGB}{0,0,196}
\definecolor{darkgreen}{RGB}{0,120,0}
\usepackage[colorlinks=true,linktocpage=true,linkcolor=darkblue,citecolor=red,urlcolor=darkblue]{hyperref}
\usepackage{cancel}
\usepackage{bbold}
\usepackage{multirow}
\usepackage{longtable}
\usepackage{color}
\usepackage[normalem]{ulem}
\usepackage{hyperref}
\usepackage{bigints}
\usepackage{xparse}
\usepackage{physics}
\usepackage{verbatim}
\usepackage{minibox}
\usepackage{comment}
\usepackage{appendix}
\usepackage{marginnote}
\usepackage{graphicx}
\usepackage[nice]{nicefrac}
\usepackage{amsmath}
\usepackage{hepunits}

\def\be{\begin{equation}}
	\def\ee{\end{equation}}
\newcommand{\bel}[1]{\begin{eqnarray}\label{#1}}
	\newcommand{\eel}{\end{eqnarray}}
\def\barr{\begin{array}}
	\def\earr{\end{array}}
\def\beq{\begin{eqnarray}}
	\def\eeq{\end{eqnarray}}
\def\bfig{\begin{figure}}
	\def\efig{\end{figure}}
\newcommand{\bea}{\begin{eqnarray}}
	\newcommand{\eea}{\end{eqnarray}}

\def\LB{\left(}
\def\RB{\right)}
\def\LSB{\left[}
\def\RSB{\right]}

\newcommand{\nn}{\nonumber}
\newcommand{\f}[2]{\frac{#1}{#2}}
\newcommand{\onehalf}{{\nicefrac{1}{2}}}

\newcommand{\p}{\partial}

\newcommand{\rf}[1]{Eq.~(\ref{#1})}

\newcommand{\rfn}[1]{(\ref{#1})}

\newcommand{\rff}[1]{Fig.~\ref{#1}}
\newcommand{\rfc}[1]{Ref.~\cite{#1}}
\newcommand{\rfcs}[1]{Refs.~\cite{#1}}

\def\a{\alpha}
\def\b{\beta}
\def\g{\gamma}

\def\LR{\left(} 
\def\RR{\right)}
\def\LS{\left[} 
\def\RS{\right]}

\newcommand{\ch}[1]{\cosh#1}

\def\half{\frac{1}{2}}

\def\GLW{{\rm GLW}}

\def\nUi{n_{(0),i}}

\def\eUi{\varepsilon_{(0),i}}

\def\PUi{P_{(0),i}}

\newcommand{\lab}[1]{\label{#1}}
\def\nn{\nonumber}

\def\av{{\boldsymbol a}}
\def\bv{{\boldsymbol b}}

\def\bfug{\xi_{\rm B}}
\def\tfug{\xi}

\def\be{\begin{equation}}
\def\ee{\end{equation}}
\def\ba{\begin{eqnarray}}
\def\ea{\end{eqnarray}}   

\def\a{\alpha}
\def\b{\beta}
\def\g{\gamma}

\def\LR{\left(} 
\def\RR{\right)}
\def\LS{\left[} 
\def\RS{\right]}

\def\half{\frac{1}{2}}

\def\GLW{{\rm GLW}}

\def\av{{\boldsymbol a}}
\def\bv{{\boldsymbol b}}

\def\epsLmnab{\epsilon_{\mu\nu\alpha\beta}}
\def\epsUmnab{\epsilon^{\mu\nu\alpha\beta}}

\def\epsLmnab{\epsilon_{\mu\nu\alpha\beta}}
\def\epsUmnab{\epsilon^{\mu\nu\alpha\beta}}

\def\epsUabgd{\epsilon^{\alpha\beta\gamma\delta}}

\def\epsLmnab{\epsilon_{\mu\nu\alpha\beta}}
\def\epsUmnab{\epsilon^{\mu\nu\alpha\beta}}

\def\epsLmnab{\epsilon_{\mu\nu\alpha\beta}}
\def\epsUmnab{\epsilon^{\mu\nu\alpha\beta}}

\def\epsUabgd{\epsilon^{\alpha\beta\gamma\delta}}

\def\half{\frac{1}{2}}
\def\GLW{{\rm GLW}}

\def\n0{n_{(0)}}
\def\e0{\varepsilon_{(0)}}
\def\P0{P_{(0)}}

\def\EM{EM }

\def\MHD{MHD }

\def\BVf{Boltzmann-Vlasov (BV) }
\def\BV{BV\,}

\def\feqpmi{f^\pm_{{\rm eq},i}}

\def\fpmi{f^\pm_i}

\newcommand{\lie}[2]{\pounds_{#1}\,#2}

\def\re{\mathrm{e}}
\def\echarge{\ensuremath{\rho_e}}
\def\cond{\ensuremath{{\sigma_e}}}

\newcommand{\GeV}{{\rm \,GeV}}
\newcommand{\MeV}{{\rm \,MeV}}
\newcommand{\fm}{\rm{\,fm}}

\begin{document}
\preprint{}

\title{Spin polarization dynamics in the Bjorken-expanding resistive MHD background}

\author{Rajeev Singh} 
\email{rajeev.singh@ifj.edu.pl}
\affiliation{Institute of Nuclear Physics Polish Academy of Sciences, PL-31-342 Krak\'ow, Poland}
\author{Masoud Shokri}
\email{mshokri@ipm.ir}
\affiliation{IPM, School of Particles and Accelerators, P.O. Box 19395-5531, Tehran, Iran}
\author{Radoslaw Ryblewski} 
\email{radoslaw.ryblewski@ifj.edu.pl}
\affiliation{Institute of Nuclear Physics Polish Academy of Sciences, PL-31-342 Krak\'ow, Poland}
\date{\today} 
%
\begin{abstract}
Evolution of spin polarization in the presence of external electric field is studied for collision energies $\sqrt{s_{\rm NN}}=27\,{\rm GeV}$ and $\sqrt{s_{\rm NN}}=200\,{\rm GeV}$. The numerical analysis is done in the perfect-fluid Bjorken-expanding resistive magnetohydrodynamic background and novel results are reported. In particular, we show that the electric field plays a significant role in the competition between expansion and dissipation.
\end{abstract}
\date{\today}

\maketitle
%
%

\section{Introduction}
\label{sec:introduction}
In recent years, relativistic hydrodynamics has become a commonly accepted tool for the description of relativistic heavy-ion collisions~\cite{Florkowski:1321594,Jaiswal:2016hex,Romatschke:2017ejr,Florkowski:2017olj,Schenke:2021mxx}, which allows us to draw a uniform picture of the complicated processes taking place in these events. It has been quite successful in describing the collective phenomena~\cite{Gale:2013da}, and hence, it is supposed to be applicable to explain the QGP (Quark-Gluon-Plasma) dynamics from the very early stages of evolution~\cite{Heller:2015dha,Shokri:2020cxa}. Despite the triumph of hydrodynamics in explaining physical observables, there are certain quantum aspects of the produced QCD matter that may not be understood using the standard formulations of relativistic hydrodynamics and indicate directions for many new investigations.
In particular, recent measurements of spin polarization of $\Lambda$ hyperons~\cite{STAR:2017ckg,Adam:2018ivw,Adam:2019srw,Acharya:2019ryw,Kornas:2019} indicate that the incorporation of spin degrees of freedom into the standard hydrodynamics framework may be necessary for understanding the spin polarization of final hadrons. The first attempt to formulate relativistic hydrodynamics with spin as a dynamic quantity has been proposed in Ref.~\cite{Florkowski:2017ruc}; for various follow-up theoretical investigations see, for instance,~\cite{Florkowski:2017dyn,Becattini:2018duy,Florkowski:2018ahw,Florkowski:2018fap,Florkowski:2019voj,Florkowski:2019qdp,Singh:2020rht,Tinti:2020gyh,Bhadury:2020puc,Bhadury:2020cop,Shi:2020htn,Florkowski:2021pkp,Gallegos:2021bzp,Liu:2021uhn,Fu:2021pok}. Other theoretical studies~\cite{Becattini:2009wh,Becattini:2013fla,Becattini:2013vja,Baznat:2015eca,Becattini:2016gvu,Karpenko:2016jyx,Xie:2017upb,Sun:2017xhx,Li:2017slc,Becattini:2017gcx,Wei:2018zfb,Xia:2018tes,Sun:2018bjl,Gao:2019znl,Ivanov:2019wzg,Kapusta:2019ktm,Gao:2020lxh,Deng:2020ygd,Serenone:2021zef,Rindori:2021quq} deal mainly with the spin polarization during the freeze-out stage of heavy-ion collisions, where they consider that the thermal vorticity is the basic hydrodynamic quantity which gives rise to spin polarization.

Grounded on basic physical arguments~\cite{Kharzeev:2007jp} and simulations~\cite{Huang:2015oca,Oliva:2019kin}, large electromagnetic (EM) fields are produced during heavy-ion collisions.
The typical scales of the initial field strength are of the order of $eE/m_\pi^2\sim eB/m_\pi^2\sim \order{1}$, with $m_\pi$ being the pion mass and $e$ denoting the elementary electric charge. If the \EM fields do not decay too quickly, they may modify different aspects of the fireball dynamics including the dynamics of spin polarization.
Although the production of large \EM fields in heavy-ion collisions is not of any doubt, their dynamics is not yet settled. Thus, recent years have observed a significant attraction to the applications of both analytical~\cite{Roy:2015kma,Pu:2016ayh,Shokri:2017xxn,Florkowski:2018ubm,Shokri:2018qcu,Shokri:2019rsc} and numerical~\cite{Inghirami:2016iru,Inghirami:2019mkc} solutions of relativistic magnetohydrodynamics (MHD) to heavy-ion collisions.

In this work, we assume the fluid in the microscopic scale to be composed of noninteracting quark-like quasi-particles of $N_f$ flavors in equilibrium, which admits a kinetic description according to the \BVf equation. By this virtue, we are not taking into account the direct coupling between the \EM fields and spin degrees of freedom.
The stationary solution of the \BV equation is obtained, in particular, in \rfc{Weickgenannt:2019dks} as the zeroth-order in $\hbar$ expansion. In this solution, modification of the chemical potential permits the electric field to exist in equilibrium~\cite{Hernandez:2017mch}. Although the electric field vectors may cancel out in the event-by-event averaging~\cite{Huang:2015oca}, the  modifications of thermodynamics that they induce, do not. We use the stationary solution to the \BV equation to derive the modification of hydrodynamic variables in equilibrium. The results are then plugged into the MHD equations, with the solutions of the Maxwell equations in the case of Bjorken flow derived in \rfc{Shokri:2017xxn}, to find the dynamics of temperature and chemical potential. We finally use the acquired background dynamics in the spin conservation law to study the spin polarization dynamics. The evolution of the hydrodynamic variables is modified both by the Joule heating (JH) term at the macroscopic level, and the modification of the thermodynamics at the microscopic level. Consequently, the competition between the JH and expansion gives rise to novel results.

\smallskip

The structure of the manuscript is as follows: We start by modifying the perfect-fluid background in the presence of the external electric field in Sec.~\ref{sec:HydroEM} and setting up the necessary hydrodynamic framework for the study of spin polarization. In Sec.~\ref{sec:spinangular} the details about the spin polarization tensor and the form of the spin tensor are given. Sec.~\ref{sec:dynamics} deals with the evolution of \EM fields and conservation laws. In Sec.~\ref{sec:numerics} we present numerical results for the thermodynamic variables and spin polarization coefficients in the perfect-fluid background. Finally, we summarize the key results and interpretations of our work and outline possible future extensions in Sec.~\ref{sec:summary}.

\smallskip

\textbf{Notation and Conventions:} In this paper, we use ``mostly minus'' metric convention. The scalar (or dot) product of two four-vectors $a^{\alpha}$ and $b^{\alpha}$ reads $a \cdot b =a^{\alpha}b_{\alpha}= g_{\alpha \beta} a^\alpha b^\beta = a^0 b^0 - \av \cdot \bv$, where three-vectors are denoted by bold font. For the  Levi-Civita tensor $\epsilon^{\alpha\beta\gamma\delta}$ we adopt the convention $\epsilon^{0123} = -\epsilon_{0123} =+1$. We denote the Lie derivative of a tensor $X$ of arbitrary rank with respect to a vector $V$ as $\lie{V}{X}$. We use a shorthand notation for antisymmetrization by a pair of square brackets. For example, for arbitrary rank-two covariant tensor $M$ we have $M_{[\mu \nu]} = \onehalf \left(M_{\mu\nu} - M_{\nu\mu} \right)$. Throughout the paper we use natural units {\it i.e.} $c = \hbar = k_B~=1$.
\section{Hydrodynamic equations in the presence of electromagnetic fields}
\label{sec:HydroEM}
The \EM fields may modify the spin hydrodynamics formalism in different ways. In the present work, we study the dynamics of spin polarization in the presence of a background gauge field $A_\mu$. We assume that each fluid element consists of quark-like quasi-particles of $N_f$ flavors, which admit a classical kinetic description. Following \rfcs{Florkowski:2018fap,Bhadury:2020puc} we assume that the single-particle distribution function for particles and antiparticles can be factorized into spin-dependent and spin-independent parts as
\be
f^\pm_{s,\rm eq}(x,p,s) = f^\pm_{\rm eq}(x,p)\exp(\half\omega_{\alpha \beta}(x) s^{\alpha\beta})\,,
\label{eq:fpm-spin}
\ee
where $f^\pm_{\rm eq}(x,p)$ is the stationary solution to the Boltzmann equation, $\omega_{\alpha \beta}$ is the spin polarization tensor (see Sec.~\ref{sec:spinangular} for discussion) and $s^{\alpha\beta}$ is the internal angular momentum~\cite{Mathisson:1937zz} for massive spin-$\onehalf$ particles defined in terms of spin four-vector $s^{\alpha}$ and particle four-momentum $p^{\alpha}$~\cite{Itzykson:1980rh} 
\bea
s^{\alpha\beta} = \f{1}{m} \epsUabgd p_\gamma s_\delta,
\label{eq:salbe}
\eea
where  $m$ is the particle mass.

In the rest of this section, we ignore possible modification of the spin-dependent part of the distribution function due to \EM field by assuming that $\omega_{\alpha \beta}$ is small~\cite{Florkowski:2018fap,Florkowski:2019qdp,Bhadury:2020cop}.
\subsection{The stationary solution to the \BVf equation} 
In the presence of \EM fields, we employ the stationary solution to the \BV equation~\cite{DeGroot:1980dk}. The relativistic \BV equation in the collisionless limit reads
\be
p^{\mu} \partial_{\mu} \fpmi \pm q_iF^{\mu \nu} p_{\nu}  \partial^p_{\mu} \fpmi  = 0\,,
\label{eq:bv}
\ee
where $i=1, \dots, N_f$ is the flavor index, $q_i$($-q_i$) is the (anti)particle electric charge for each flavor and $F_{\mu\nu}=\p_\mu A_\nu-\p_\nu A_\mu$ is the \EM field strength tensor. All quarks have the same baryon number while their electric charges differ. In the global equilibrium, the solution to \rf{eq:bv} reads~\cite{DeGroot:1980dk,Weickgenannt:2019dks}
\be
\feqpmi (x,p) = \exp\Big[\pm\frac{\bfug}{3}-\beta^\mu\left(p_\mu\pm q_i A_\mu\right)\Big]\,,
\label{eq:bv-sol}
\ee
where $\bfug$ is the ratio of baryon chemical potential $\mu$ over temperature $T$, $\bfug \equiv \mu/T$, and $\beta^\mu$ is the ratio of fluid flow vector $U^\mu$ and temperature, $\beta^\mu \equiv U^\mu/T$. Plugging the solution \eqref{eq:bv-sol} into \rf{eq:bv} gives rise to 
\be
\half p^\mu p^\nu \lie{\beta}{g_{\mu\nu}} \pm q_i p^\mu \lie{\beta}{A_\mu} = 0\,.
\label{eq:bv-equil}
\ee
Here $\lie{\beta}{X}$ is the Lie derivative of a tensor with respect to $\beta$ and in particular~\cite{parker_toms_2009}
\ba
\label{eq:liederivative}
\lie{\beta}{A_\mu} &=& \b^\nu\p_\nu A_\mu + A_\nu \p_\mu \b^\nu,\\
\lie{\beta}{g_{\mu\nu}} &=& \p_\mu \b_\nu + \p_\nu \b_\mu\,.
\ea
Equation~\eqref{eq:bv-equil} is satisfied in global equilibrium for which we have 
\be
\lie{\beta}{g_{\mu\nu}} = 0 \,,\quad \lie{\beta}{A_\mu} = 0\,.
\label{eq:global-eq}
\ee
We rewrite the latter relation above using Eq. \eqref{eq:liederivative} as follows
\begin{eqnarray}
	\lie{\beta}{A_\nu} &=& \b^\mu\left(\p_\mu A_\nu-\p_\nu A_\mu\right) + \b^\mu \p_\nu A_\mu + A_\mu \p_\nu \b^\mu\nn\\
	&=& \b^\mu F_{\mu\nu} + \p_\nu \left(\b \cdot A\right)=0\,.
	\label{eq:gauge-field-lie}
\end{eqnarray}
The Faraday tensor $F_{\mu\nu}$ can be decomposed with respect to the four-velocity $U_\mu$ in the following way~\cite{Bekenstein-MHD}
\be
F_{\mu\nu} = E_\mu U_\nu -E_\nu U_\mu + \epsLmnab ~U^\a B^\b\,,
\label{eq:f-decomp}
\ee
where the \EM four-vectors are defined as
\be
E^\mu \equiv F^{\mu\nu} U_\nu\,,\quad B^\mu \equiv \half \epsUmnab F_{\nu\a} U_\b\,.
\label{eq:em-fv}
\ee
Plugging \rf{eq:f-decomp} into \rf{eq:gauge-field-lie} and expressing $U^\mu$ with $\beta^\mu$ gives rise to
\ba
\frac{E_\mu}{T} = \p_\mu\left(\b \cdot A\right)\, .
\label{eq:Emu}
\ea
Integrating \rf{eq:Emu} and assuming that $E_\mu$ and $T$ are slowly varying at the microscopic scale leads to
\be
\beta\cdot A = \f{E_\mu}{T}\int d x^\mu\,,
\ee
up to a gauge transformation that should be absorbed into the quark baryon chemical potential $\mu/3$~\cite{Kovtun:2016lfw}.
By this virtue, the solution \eqref{eq:bv-sol} can be rewritten as
\be
\feqpmi (x,p) = \exp\left(\pm\tfug_i-\beta^\mu p_\mu\right)\,,
\label{eq:eq-dist-function}
\ee
where,
\be
\tfug_i = \bfug - q_i  \f{E_\mu}{T}\int d x^\mu.
\label{eq:xii}
\ee
The distribution function \eqref{eq:eq-dist-function} has an important implication. Even if the event-by-event average of the electric field vanishes~\cite{Huang:2015oca}, its fingerprint in the distribution function may survive.

Thus, the spin distribution function Eq. \eqref{eq:fpm-spin}, can be written using Eq. \eqref{eq:eq-dist-function} in the small polarization limit as
\ba
f^\pm_{i,s,\rm eq}(x,p,s) = \feqpmi(x,p)\Big[1+\half\omega_{\alpha \beta}(x) s^{\alpha\beta}\Big]\,.
\label{eq:fspin}
\ea
\subsection{Baryon and electric charges}\label{sec:charges}
In contrast to \rfc{Florkowski:2019qdp}, the fluid considered in this work has two different charge currents: a baryon charge current $N^\alpha$ and an electric charge current $J^\alpha$. In equilibrium we have~\cite{Bhadury:2020cop}
\ba
N^\a_{\rm eq} 
= \sum_{i}^{N_f}\!\int \! \mathrm{dP}~\mathrm{dS} \, \, p^\a \, \left[f^+_{i,s, \rm eq} \!-\!f^-_{i,s, \rm eq}  \right], 
\label{eq:Neq}
\ea
where the invariant momentum integration measure $\mathrm{dP}$ and spin integration measure $\mathrm{dS}$ is~\cite{Florkowski:2018fap} 
\bea
\mathrm{dP} = \frac{d^3p}{(2 \pi )^3 E_p}, \quad \mathrm{dS} = \frac{m}{\pi {\mathfrak{s}}} \, \mathrm{d}^4s~\delta(s\cdot s + {{\mathfrak{s}}}^2)~\delta(p\cdot s),\nn\\
\label{eq:dP}
\eea
and ${{\mathfrak{s}}}^2=3/4$ is the length of the spin vector.

Plugging the distribution function \eqref{eq:fspin} in  \rf{eq:Neq}, and keeping the terms up to first order for small polarization in $\omega_{\mu\nu}$ gives rise to~\cite{Florkowski:1321594,Florkowski:2017ruc}
\ba
N^\alpha_{\rm eq} = n\,U^\alpha  = \sum_{i}^{N_f}n_i\,U^\alpha,
\label{eq:Nmu}
\ea
where $n_i = 4 \, \sinh(\tfug_i)\, \nUi(T)$ and $\nUi(T)$ denotes the number density of spinless and neutral massive Boltzmann particles of the form~\cite{Florkowski:1321594,Florkowski:2017ruc}
\beq
\nUi(T) &=&  \f{1}{2\pi^2}\, T^3 \, \hat{m}_i^2 K_2\left(\hat{m}_i\right), \label{eq:polden}
\eeq
with $\hat{m}_i\equiv m_i/T$ being the ratio of the $i$-th flavor mass over temperature and $K_{n}\left(\hat{m}_i\right)$ is the $n$-th modified Bessel function of the second kind.

Baryon charge conservation law has the form
\ba
  \p_\alpha N^\alpha_{\rm eq}(x)  = 0,
  \label{eq:Ncon}
\ea
which is guaranteed by Eq.~\eqref{eq:bv} and it also implies that the baryon charge current is independently conserved for each flavor, namely,
\ba
\p_\alpha N_{i,{\rm eq}}^\alpha(x)  = 0,\qq{with} N_{i,{\rm eq}}^\alpha \equiv n_i U^\a\,.
\label{eq:Nconf}
\ea
Next we turn to the electric current. First, note that the conservation of electric current is an implication of the inhomogeneous Maxwell equation, namely,
\be
\p_\mu F^{\mu\nu} = J^\nu = \echarge U^\nu + \Delta^{\nu\rho}J_\rho\,,
\label{eq:ihom-maxwell}
\ee
where $\Delta^{\nu\rho}$ is the spatial projection operator expressed as 
\be
\Delta^{\nu\rho} = g^{\nu\rho} - U^\nu U^\rho\,,
\ee
and $\echarge$ is the local electric charge density given by
\ba
\echarge 
= \sum_{i}^{N_f}\!\int \! \mathrm{dP}~\mathrm{dS} \, \, \left(p\cdot U\right) \, q_i\left[f^+_{i,s, \rm eq}\!-\!f^-_{i,s, \rm eq} \right]\, 
\label{eq:rho}
\ea
which similar to \rf{eq:Nmu}, gives rise to
\be
\echarge = \sum_{i}^{N_f} q_i n_i\,.
\ee
In what follows, we use Bjorken-expanding resistive MHD in which the symmetries require electric neutrality~\cite{Shokri:2017xxn}. The fluid can have a net baryon charge density $n$ with a vanishing electric charge density $\echarge$. Such a setup is simplified but not unrealistic in the later stages of the fireball evolution. Out of equilibrium, the electric current is expressed as
\be
J^\mu = \cond E^\mu\,,
\label{eq:electric-current}
\ee
where $\cond$ is the electric conductivity and the above current is dissipative.
\subsection{Conservation of energy and linear momentum}
In equilibrium, the energy-momentum tensor of the fluid reads
\ba
T^{\mu\nu}_{\rm eq} 
= \sum_{i}^{N_f}\!\int \! \mathrm{dP}~\mathrm{dS} \, \, p^\mu p^\nu\, \left[f^+_{i,s, \rm eq}\!+\!f^-_{i,s, \rm eq}  \right]\,.
\label{eq:Tmunu}
\ea
Using the equation \eqref{eq:fspin} in above equation, we obtain
\ba
\p_\mu T^{\mu\nu}_{\rm eq} = F^{\nu\rho}J_{\rho,\rm eq}\,.
\label{eq:TconEQ}
\ea
However, the above form of the energy-momentum conservation law is implied by the diffeomorphism and gauge invariance~\cite{Jensen:2012jh} and is therefore independent of the underlying microscopic theory. By this virtue, the conservation law of energy and linear momentum has the form
\ba
\p_\a T^{\a\b}_{\rm fluid}(x) = F^{\b\g}J_{\g},
\label{eq:Tcon}
\ea
where for the perfect-fluid, the energy-momentum tensor $T^{\a\b}_{\rm eq}$ is expressed as
\ba
T^{\a\b}_{\rm eq} &=&  (\varepsilon + P ) U^\a U^\b - P g^{\a\b}\,,
\label{Tmn}
\ea
with the energy density and pressure having the form
\ba
\varepsilon = 4\sum_{i}^{N_f}\cosh(\tfug_i) \, \eUi(T),
\label{enden}
\ea
and 
\ba
P = 4\sum_{i}^{N_f}\cosh(\tfug_i) \, \PUi(T),
\label{prs}
\ea
respectively.

Similar to  (\ref{eq:polden}), $\eUi(T)$ and $\PUi(T)$ are the energy density and pressure for spinless and neutral massive Boltzmann particles defined  as~\cite{Florkowski:1321594,Florkowski:2017ruc}
\beq
\eUi(T) &=& \f{1}{2\pi^2} \, T^4 \, \hat{m}_i ^2
 \Big[\hat{m}_i  K_{1} \left( \hat{m}_i  \right)+3 K_{2}\left(\hat{m}_i\right) \Big],  \label{eneden} 
\eeq
and
\beq 
\PUi(T) &=& T \, \nUi(T)  , \label{P0}
\eeq
respectively.
\subsection{Entropy conservation}
At this stage, we would like to comment on entropy conservation in the presence of background electric fields. The entropy current reads~\cite{Bhadury:2020cop}
\bea
H^\mu &=& -\sum_{i}^{N_f}\!\!\int \! \mathrm{dP}~\mathrm{dS} \, p^\mu
\LSB 
f^+_{i,s, \rm eq}\LB \log f^+_{i,s, \rm eq} -1 \RB 
\right. \nn \\ && \left. \hspace{1.5cm}
+ 
f^-_{i,s, \rm eq} \LB \log f^-_{i,s, \rm eq} -1 \RB \RSB.
\label{eq:H1}
\eea 
Plugging \rf{eq:fpm-spin} with \rf{eq:bv-sol} into the above equation, we obtain~\cite{Bhadury:2020cop}
\bea
H^\mu = P\beta^\mu + \beta_\alpha T^{\mu \alpha}_{\rm eq} -\frac{1}{2} \omega_{\alpha \beta} S^{\mu,\alpha \beta}_{\rm eq}
-\sum_{i}^{N_f}\xi_i(x) N^\mu_
{i,\rm eq}\,\nn\\
\label{eq:H2}
\eea 
where $S^{\mu,\alpha \beta}_{\rm eq}$ is the spin tensor.

In global equilibrium, where \rf{eq:global-eq} holds one has
\ba
\p_\mu \left(P\beta^\mu + \beta_\alpha T^{\mu \alpha}_{\rm eq}\right) &=& \beta_\alpha \p_\mu T^{\mu \alpha}_{\rm eq} 
\nn\\
&&\hspace{-2cm}=\beta_\alpha F^{\a\b}J_{\b,\rm eq}
= - \sum_{i}^{N_f} q_i E_\mu N^\mu_	{i,\rm eq}\,,
\ea
wherein \rf{eq:TconEQ} was used.
Using the above relation, and \rf{eq:Nconf} in the divergence of \eqref{eq:H2} gives rise to
\ba
\p_\mu H^\mu &=& - \sum_{i}^{N_f} N^\mu_
{i,\rm eq} \p_\mu\left(\xi_i(x)+q_i E_\mu\right)
\nn\\
&=&-nT \beta^\mu\p_\mu \bfug = -nT \lie{\beta}{\bfug} = 0\,.
\label{eq:dmuHmu}
\ea
We conclude that the electric field does not induce entropy production in global equilibrium provided the chemical potential is modified according to \rf{eq:xii}.

We also point out that the entropy current has the contribution from the polarization. However, since there is no direct coupling between spin and electromagnetic fields in our setup, we have neglected such terms as they were already taken care of in Ref.~\cite{Bhadury:2020cop}. We should note that the entropy current analysis does not rely on any approximation, and the Eq.~\eqref{eq:H2} is exact. Eq.~(\ref{eq:dmuHmu}) and the analysis given in Ref.~\cite{Bhadury:2020cop} admits that the entropy is conserved in, and only in, equilibrium.

The setup that is worked out in this section is called the strong electric field regime by certain authors~\cite{Hernandez:2017mch}. It should be emphasized that in the rest of the current work we consider resistive \MHD   equations with the electrical conductivity as the only source of dissipation. Out of equilibrium, the entropy production has the form
\be
\p_\mu H^\mu = \frac{\cond}{T}E^2\,,
\ee 
where $E\equiv\sqrt{-E^\mu E_\mu}$.
\section{Spin polarization tensor and conservation of spin angular momentum}
\label{sec:spinangular}
\subsection{Spin polarization tensor}
The spin polarization tensor $\omega_{\mu\nu}$ is an antisymmetric rank-two tensor which can be always expressed as follows
\beq
\omega_{\mu\nu} &=& \kappa_\mu U_\nu - \kappa_\nu U_\mu + \epsilon_{\mu\nu\a\b} U^\a \omega^{\b}, \lab{spinpol1}
\eeq
where $U^\mu$ is the fluid four-velocity and $\kappa^\mu$ and $\omega^\mu$ are yet another four-vectors~\cite{Florkowski:2017ruc}. Any part of the $\kappa^{\mu}$ and $\omega^{\mu}$  parallel to $U^{\mu}$ does not contribute to the right-hand side of~Eq.~(\ref{spinpol1}).
Therefore, we assume that $\kappa^{\mu}$ and $\omega^{\mu}$ fulfill the following orthogonality conditions
\beq
\kappa\cdot U = 0, \quad \omega \cdot U = 0  \lab{ko_ortho}.
\eeq
Hence $\kappa_\mu$ and $\omega_\mu$ can be written as
\beq
\kappa_\mu &=& \omega_{\mu\a} U^\a \equiv a_X X_\mu + a_Y Y_\mu + a_Z Z_\mu \nn\\
\omega_\mu &=& \half \epsilon_{\mu\a\b\g} \omega^{\a\b} U^\g \equiv b_X X_\mu + b_Y Y_\mu + b_Z Z_\mu \lab{eq:kappaomega}
\eeq
Here the scalar quantities $a_{X}, a_{Y}, a_{Z},  b_{X}, b_{Y},$ and $ b_{Z}$ are called spin polarization coefficients.

A general expression for the spin polarization tensor in terms of $\kappa$ and $\omega$ four-vectors has the form~\cite{Florkowski:2019qdp},
\beq
\omega_{\mu\nu} &=& a_X (X_\mu U_\nu - X_\nu U_\mu)+ a_Y (Y_\mu U_\nu - Y_\nu U_\mu)\nn \\
&& + a_Z (Z_\mu U_\nu - Z_\nu U_\mu) \nonumber \\
&& + \, \epsilon_{\mu\nu\alpha\beta} U^\alpha (b_X X^\beta + b_Y Y^\beta + b_Z Z^\beta), \label{eq:omegamunu}
\eeq
where $X$, $Y$ and $Z$ together with $U$ form a four-vector basis satisfying the following normalization conditions
\begin{eqnarray}
 && \,\,U \cdot U = 1\label{UU}\\
X \cdot X \,\,&=& \,\, Y \cdot Y \,\,=\,\, Z \cdot Z \,\,=\,\, -1, \\ \label{XXYYZZ}
X \cdot U\,\, &=& \,\,Y \cdot U \,\,=\,\, Z \cdot U \,\,=\,\, 0,  \\ \label{XYZU}
X \cdot Y\,\, &=& \,\, Y \cdot Z \,\,=\,\, Z \cdot X \,\,=\,\, 0.  \label{XYYZZX}
\end{eqnarray}
\subsection{Conservation of angular momentum}
In the formalism by de Groot, van
Leeuwen, and van Weert (GLW)  the energy-momentum tensor is symmetric, hence, the angular momentum conservation implies the conservation of the spin tensor and therefore we can write~\cite{Florkowski:2018ahw}
\beq
\p_\a S^{\a , \beta \gamma }_{\rm GLW}(x)&=& 0,
\label{eq:SGLWcon}
\eeq
where in the leading-order spin polarization tensor, the GLW spin tensor is written as \cite{Florkowski:2017dyn,Florkowski:2018ahw}
\beq
S^{\a, \b\g}_\GLW 
&=& \sum_{i}^{N_f}\ch(\tfug_i)\Bigg[{\cal A}_{1,i} (U^{[\b} \omega^{\g]\a} + g^{\a[\b} \kappa^{\g]})\nn \\
&& + {\cal A}_{2,i}~U^\a U^{[\b} \kappa^{\g]} + {\cal A}_{3,i}~U^\a \omega^{\b\g}\Bigg] ,
\lab{SGLW2}
\eeq
with thermodynamic coefficients having forms
\beq
{\cal A}_{1,i} &=& -\frac{2}{\hat{m}_i^2} \frac{\varepsilon_{(0),i}(T)+P_{(0),i}(T)}{T}\label{A1},\\
{\cal A}_{2,i} &=&   2 n_{(0),i}(T)-6{\cal A}_{1,i}  \label{A2} , \\
{\cal A}_{3,i} &=&   \nUi -  {\cal A}_{1,i}  \label{A3}.
\eeq
\section{Perfect-fluid and spin dynamics}
\label{sec:dynamics}
Based on the discussion given in the previous sections in the following we will study perfect-fluid dynamics in the presence of electromagnetic fields, as given by \rf{eq:Ncon} and \rf{eq:Tcon}, and, on top of it, we consider spin evolution equations, as determined by \rf{eq:SGLWcon}.
\subsection{Boost invariant form of conservation laws}
Using \rf{eq:Nmu}, the conservation law for charge \rf{eq:Ncon} can be cast into the following form
\beq
U^{\a}\p_{\a}n+n\p_{\a}U^{\a}\equiv\frac{dn}{d\tau}+\frac{n}{\tau}=0.
\label{eq:ccons}
\eeq
Projecting \rf{eq:Tcon} on $U_{\b}$ and then using \rf{Tmn}, we also obtain
\ba
U^{\a}\p_{\a}\varepsilon+\left(\varepsilon+P\right)\p_{\a}U^{\a}&=&\cond\,E^2\,,\nn\\
\frac{d\varepsilon}{d\tau}+\frac{\varepsilon+P}{\tau}&=&\cond\,E^2.
\label{eq:encons}
\ea
Using Eqs.~(\ref{eq:omegamunu}) and \rfn{SGLW2} in \rf{eq:SGLWcon}, and contracting the resulting tensor equation with $U_\b X_\g$, $U_\b Y_\g$, $U_\b Z_\g$,  $Y_\b Z_\g$, $X_\b Z_\g$ and $X_\b Y_\g$ we obtain the following evolution equations for the spin polarization coefficients $\boldsymbol{a} = \LR a_{X}, a_{Y}, a_{Z},  b_{X}, b_{Y}, b_{Z} \RR$,~\cite{Florkowski:2019qdp},
\begin{equation}
{\rm diag}\LR
\cal{L}, \cal{L}, \cal{L}, \cal{P}, \cal{P}, \cal{P}\RR \,\,
\boldsymbol{\Dot{a}} ={\rm diag}\LR
{\cal{Q}}, {\cal{Q}}, {\cal{Q}}_1, {\cal{R}}, {\cal{R}}, {\cal{R}}_1 \RR\,\,
\boldsymbol{a}, \label{cs}
\end{equation}
respectively, where $\dot{(\dots)} \equiv U \cdot \p = \p_\tau$
and,
\beq
{\cal L}(\tau)&=&{\cal A}_{1,i},\nn\\
{\cal P}(\tau)&=&{\cal A}_{3,i},\nn\\
{\cal{Q}}(\tau)&=&-\left[\dot{{\cal L}}+\frac{3{\cal L}}{2\tau} \right],\nn\\
 {\cal{Q}}_1(\tau)&=&\left[{\cal{Q}}+\frac{{\cal L}}{2\tau}   \right],\nn\\
  {\cal{R}}(\tau)&=&-\left[\Dot{\cal P}+\frac{1}{\tau}\left({\cal P} -\frac{{\cal L}}{2} \right) \right],\nn\\
 {\cal{R}}_1(\tau)&=&\left[{\cal{R}} -\frac{{\cal L}}{2\tau}\right],
 \label{LPQR}
 \eeq
 c.f. Eq.~(\ref{eq:omegamunu}).
%
\subsection{Evolution of the \EM fields}\label{sec:em-fields}
To obtain the evolution of EM fields, one needs to simultaneously solve the Maxwell equations and the conservation laws for energy and linear momentum. In general, this is not an easy task. However, here we adopt the nonrotating or maximally boost invariant solution presented in \rfc{Shokri:2017xxn} which is given by
\be
B^\mu = B_0 \frac{\tau_0}{\tau} Y^\mu\,,\quad E^\mu = \ell E_0 \frac{\tau_0}{\tau}\re^{-\cond\left(\tau-\tau_0\right)}Y^\mu\,, 
\label{eq:em-fields} 
\ee
where $B_0$ and $E_0$ are the initial values of the magnetic and electric field at the initial proper time $\tau_0$, respectively, whereas the parameter $\ell\equiv\f{\boldsymbol{B}\cdot\boldsymbol{E}}{BE}=\pm 1$, corresponds to the parallel and anti-parallel field configurations. The solution to the resistive MHD equations mentioned above is found as follows. The translational symmetries of the Bjorken flow do not permit the magnetic field to exist in the longitudinal ($z$) direction~\cite{Shokri:2018qcu}, as well as, boost invariance requires both $E_z$ and electric charge density $\rho_e$ to vanish. Thus, both electric and magnetic fields are constrained to exist only in the transverse, i.e. $x-y$, plane. To preserve the Bjorken flow, the acceleration due to the Poynting vector must vanish, which implies that the electric and magnetic fields are either parallel or antiparallel to each other. The angle between the fields is either $0~(\ell=1)$ or $\pi~(\ell=-1)$, which  remains fixed during the evolution and is a part of the initial conditions. If one assumes that the direction of the fields is also boost invariant, the solution \eqref{eq:em-fields} is found from the Maxwell's equations. We should also emphasize that the magnetic field exists in the system, however it does not play any direct role in our setup.
Plugging \rf{eq:em-fields} into \rf{eq:xii}, gives rise to
\be
\tfug_i = \bfug - \ell\,q_i\,R_{\rm RMS} \frac{E_0}{T}\left(\frac{\tau_0}{\tau}\right)\re^{-\cond\left(\tau-\tau_0\right)}\,,
\label{eq:tfug} 
\ee
where $R_{\rm RMS}$ is the nucleon root-mean-square charge radius. The solution \eqref{eq:em-fields} implies that
\ba
\echarge = 0\,.
\ea
We note here that in our setup the neutrality is not automatically satisfied at the initial time. To see this, consider $\rho_e$ at the initial time for $N_f=3$. The neutrality puts a constraint on the initial number densities
\ba
2n_{u}(\tau_0) - n_{d}(\tau_0) - n_{s}(\tau_0) = 0\,,
\ea
As it turns out, such a constraint cannot be satisfied with the physical parameters introduced here, and there is no reason to believe that it should be satisfied by the fluid in realistic situations. On the other hand, as the fluid starts to evolve the local electric charge density relaxes to negligible values in a fraction of a Fermi, and, to a very good approximation, neutrality is reached. 
\section{Numerical results}
\label{sec:numerics}
In this section we present numerical results for the hydrodynamic variables obtained by solving Eqs.~\eqref{eq:ccons} and  \eqref{eq:encons} in the presence of external electric field. These results are then used to solve Eqs.~\eqref{cs} for the spin polarization coefficients. The calculations are performed for two choices of collision energies, i.e,  $\sqrt{s_{\rm NN}}=27\GeV$ and $\sqrt{s_{\rm NN}}=200\GeV$. The initial values for the temperature and baryon chemical potential for each collision energy read as~\cite{Karpenko:2015xea}
\begin{align*}
T_0 &= 300 \MeV\,,\quad \mu_{0} = 300\MeV \qq{for} \sqrt{s_{\rm NN}}=27\GeV\,,\\
T_0 &= 600\MeV\,,\quad \mu_{0} = 50\MeV \qq{for} \sqrt{s_{\rm NN}}=200\GeV\,.
\end{align*}
We use the constituent quark masses  at $\Lambda$'s mass scale for number of quark flavors $N_f=3$ ~\cite{BorkaJovanovic:2010yc}
\be
m_u =m_d= 0.382\GeV\,,\quad m_s = 0.537 \GeV\,.
\ee
\begin{figure}[t]
\begin{center}
\includegraphics[width=8.5cm]{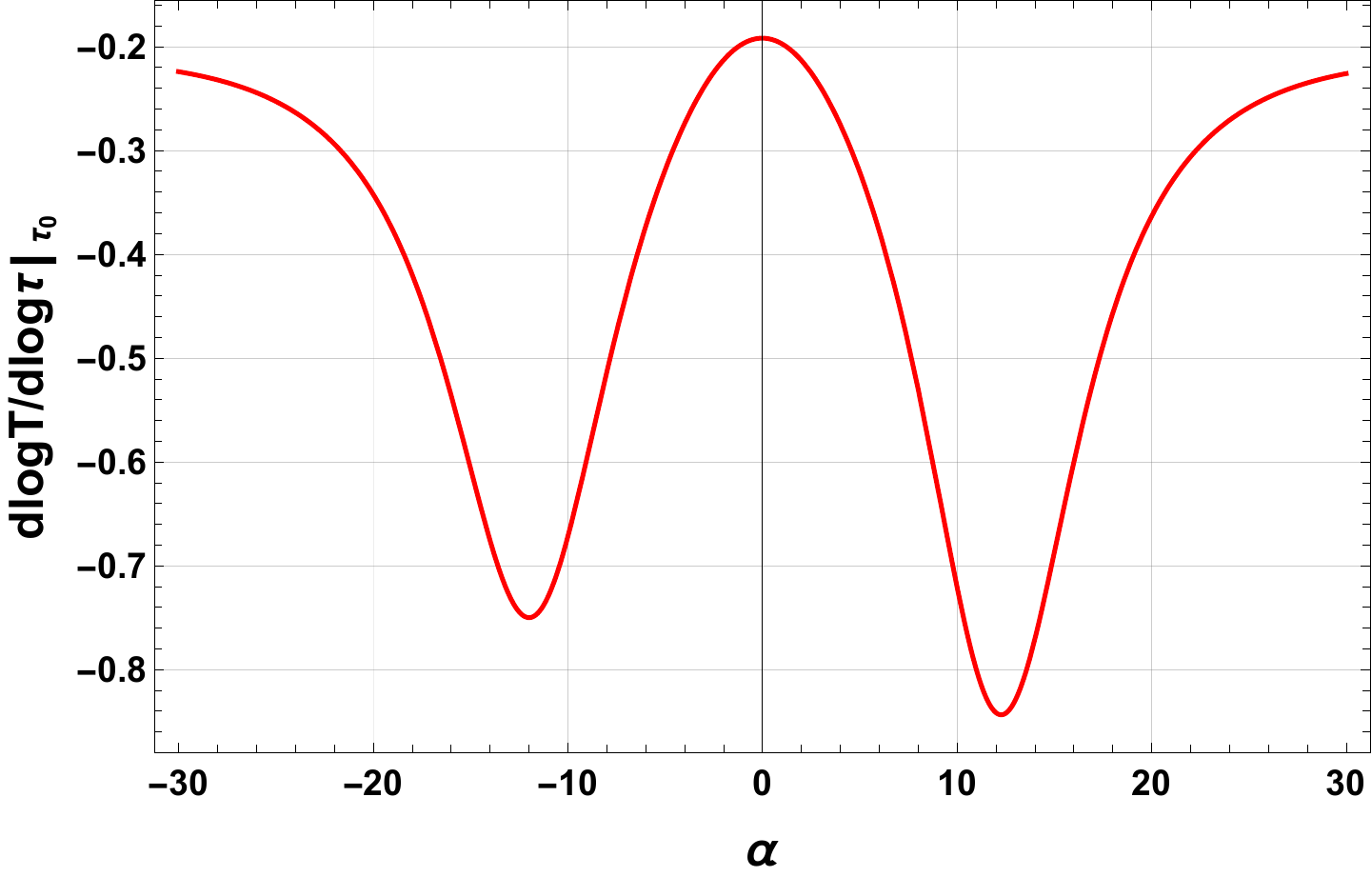}
\includegraphics[width=8.5cm]{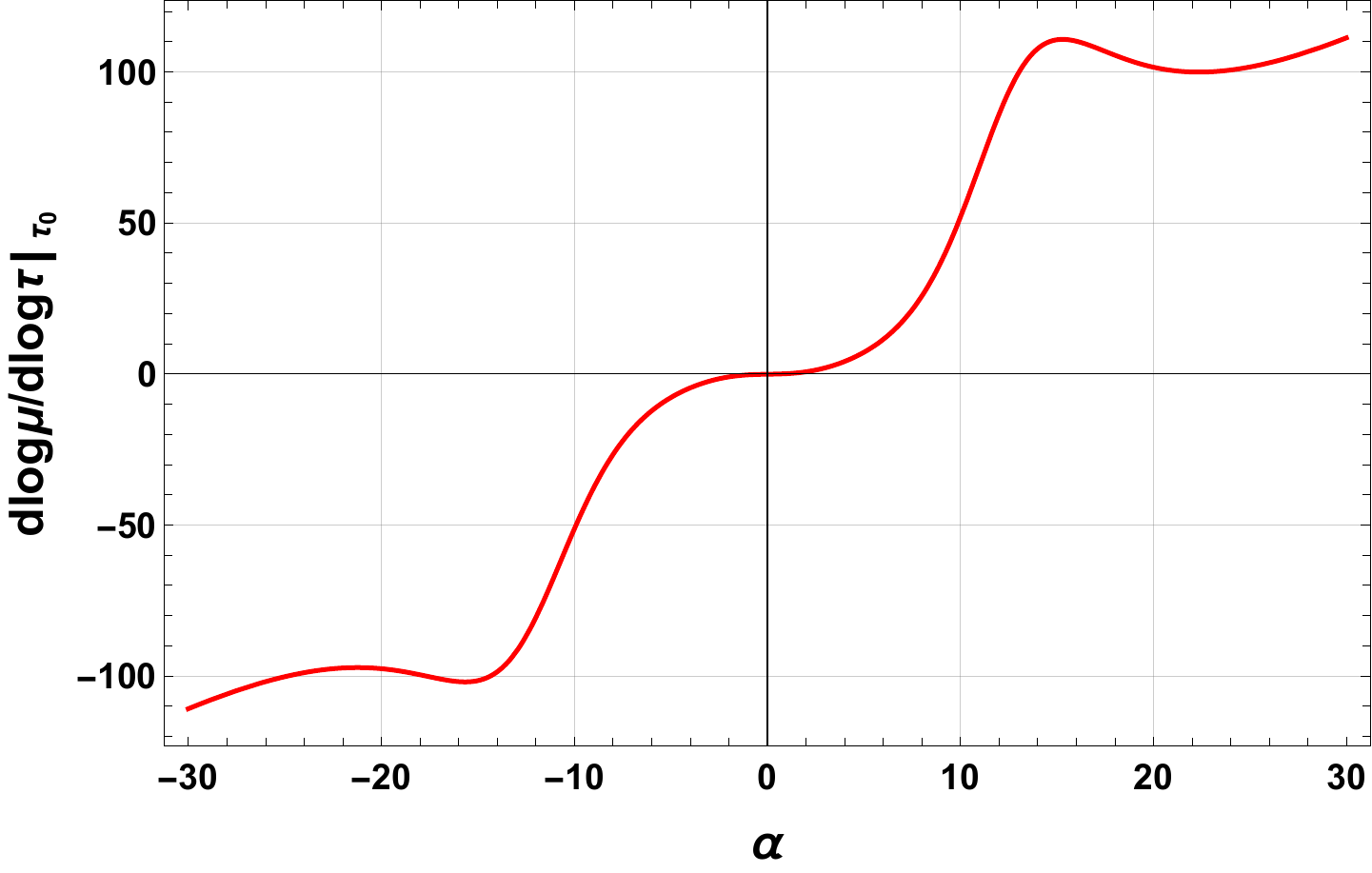}
\end{center}
\caption{\small The quantity
$(d\log T)/(d\log\tau)$ (upper panel) and
$(d\log \mu)/(d\log\tau)$ (lower panel) as a function of $\alpha$ at $\tau=\tau_{0}$ for the $\sqrt{s_{\rm NN}}=200\GeV$.}
\label{fig:derivatives}
\end{figure}
\begin{figure}[t]
\begin{center}
\includegraphics[width=8.5cm]{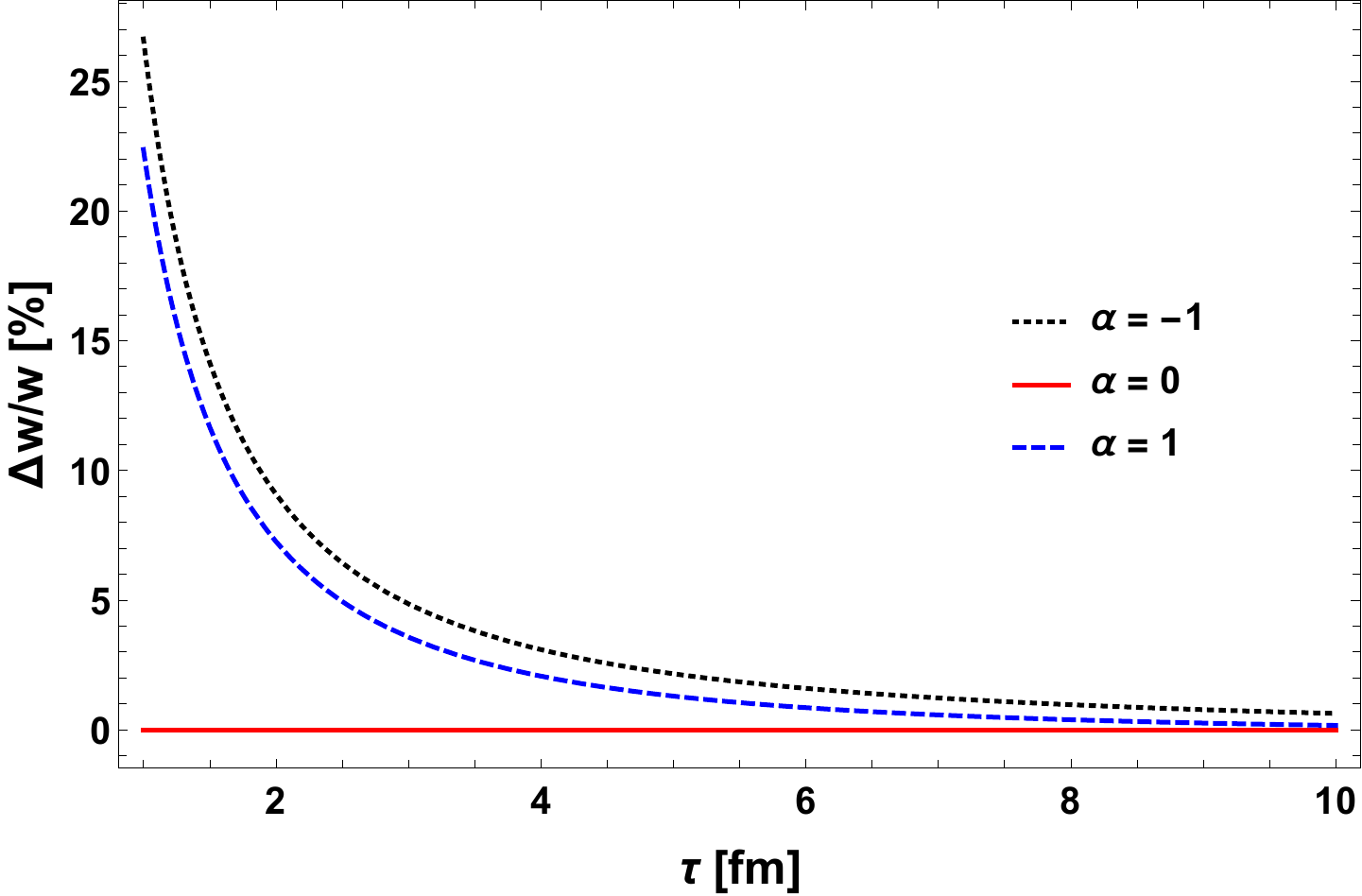}
\end{center}
\caption{\small Percentage of enthalpy density enhancement with respect to $\alpha=0$ case: $\Delta w/w \equiv \left[(\varepsilon+P)_{\alpha\neq 0}-(\varepsilon+P)_{\alpha= 0}\right]/(\varepsilon+P)_{\alpha= 0}$. 
The figure is for the $\sqrt{s_{\rm NN}}=27\GeV$ case.}
\label{fig:hplot}
\end{figure}
The initial proper time for both energies is chosen to be $\tau_0=1\fm$, and we adopt $R_{\rm RMS}=4.3\fm$ from \rfc{Gubser:2008pc}. The electric conductivity to second order approximation in $\mu/T$ is given by~\cite{Aarts:2020dda}
\ba
\cond (T, \mu) =0.37~Q_{e}~T \LS 1+0.15~\LR\frac{\mu}{T}\RR^2\RS,
\label{eq:conductivity}
\ea
where $Q_{e}=(2/3) e^2$ is the sum over flavors of the quark electric charges squared. To employ different values for the initial electric field $E_0$, we introduce the following parameter 
\be
\a \equiv \ell~\frac{e E_0}{m_\pi^2}\,,
\ee
where we choose $eE_0/m_\pi^2 = 1$ for $\sqrt{s_{\rm NN}}=200\GeV$ and scale it down linearly with $\sqrt{s_{\rm NN}}$ for $\sqrt{s_{\rm NN}}=27\GeV$~\cite{Huang:2015oca}. As a result, $\a$ is our only free parameter for each chosen collision energy. To understand the dynamics of $T$ and $\mu$, it is suitable to rewrite \rf{eq:encons} in the following form
\beq
\tau \frac{{d\varepsilon}}{{d\tau}} = - w + \cond\tau E^2\,,
\label{eq:competition}
\eeq
where, $w=\varepsilon+P$ is the enthalpy density.
\begin{figure}[t]
\begin{center}
\includegraphics[width=8.5cm]{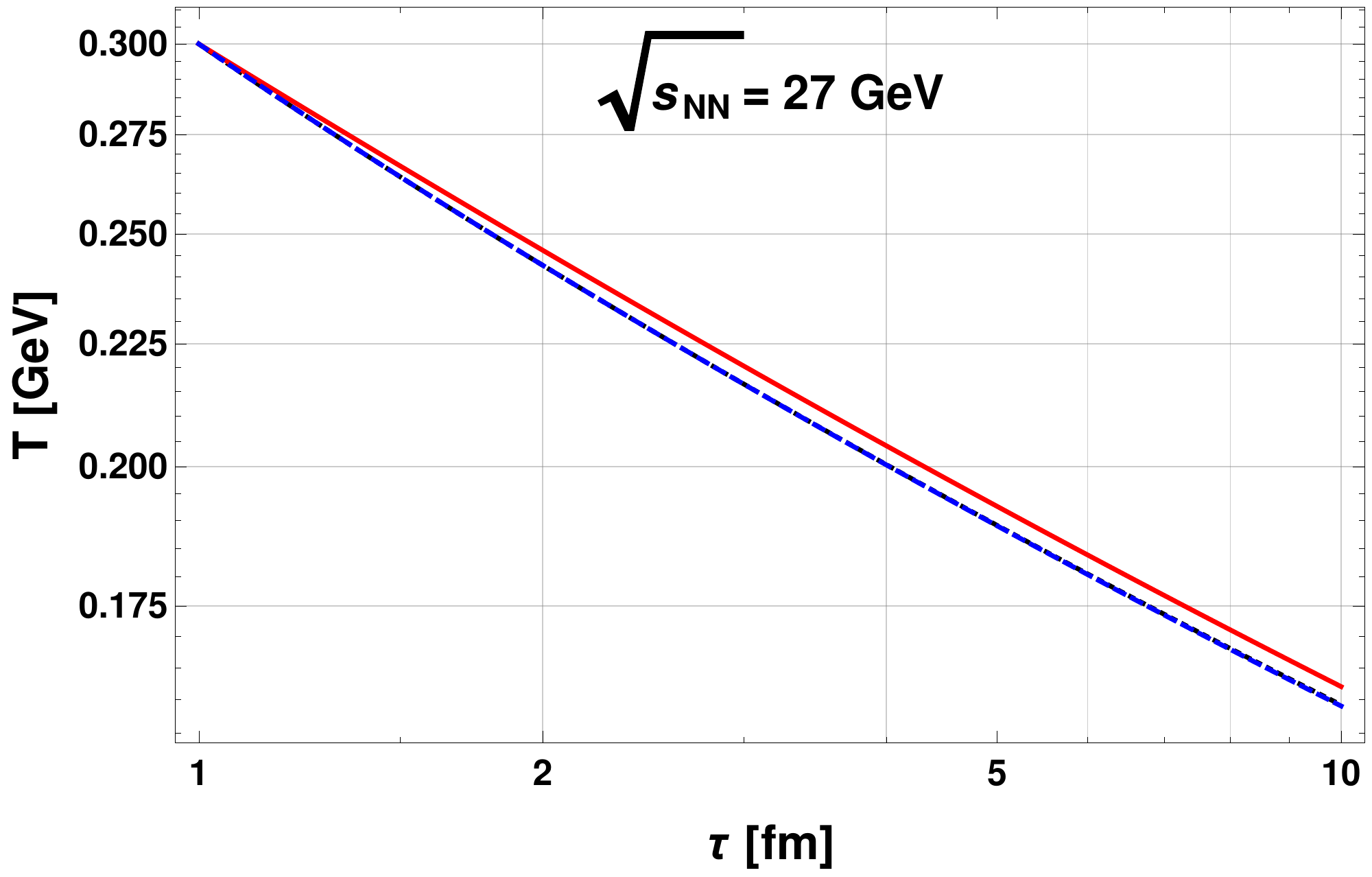}
\includegraphics[width=8.5cm]{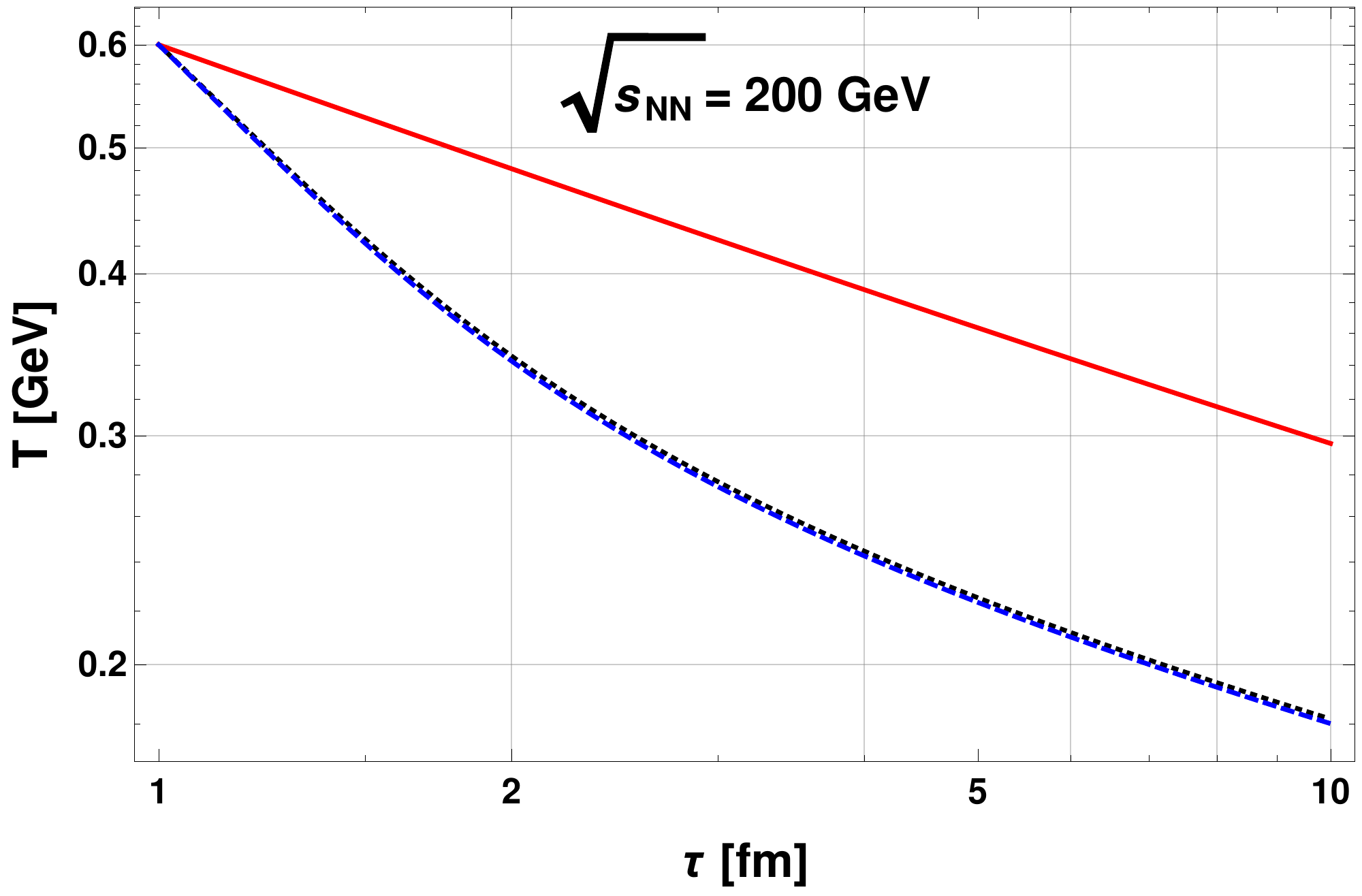}
\end{center}
\caption{\small Temperature profile with initial temperature $T_0 = 300 \MeV$ for $\sqrt{s_{\rm NN}}=27\,{\rm GeV}$ (upper panel) and $T_0 = 600 \MeV$ for $\sqrt{s_{\rm NN}}=200\,{\rm GeV}$ (lower panel). Dotted black line is for $\alpha = -8$, red line is for $\alpha=0$ and dashed blue line is for $\alpha=8$.}
\label{fig:T}
\end{figure}
\begin{figure}[t]
\begin{center}
\includegraphics[width=8.5cm]{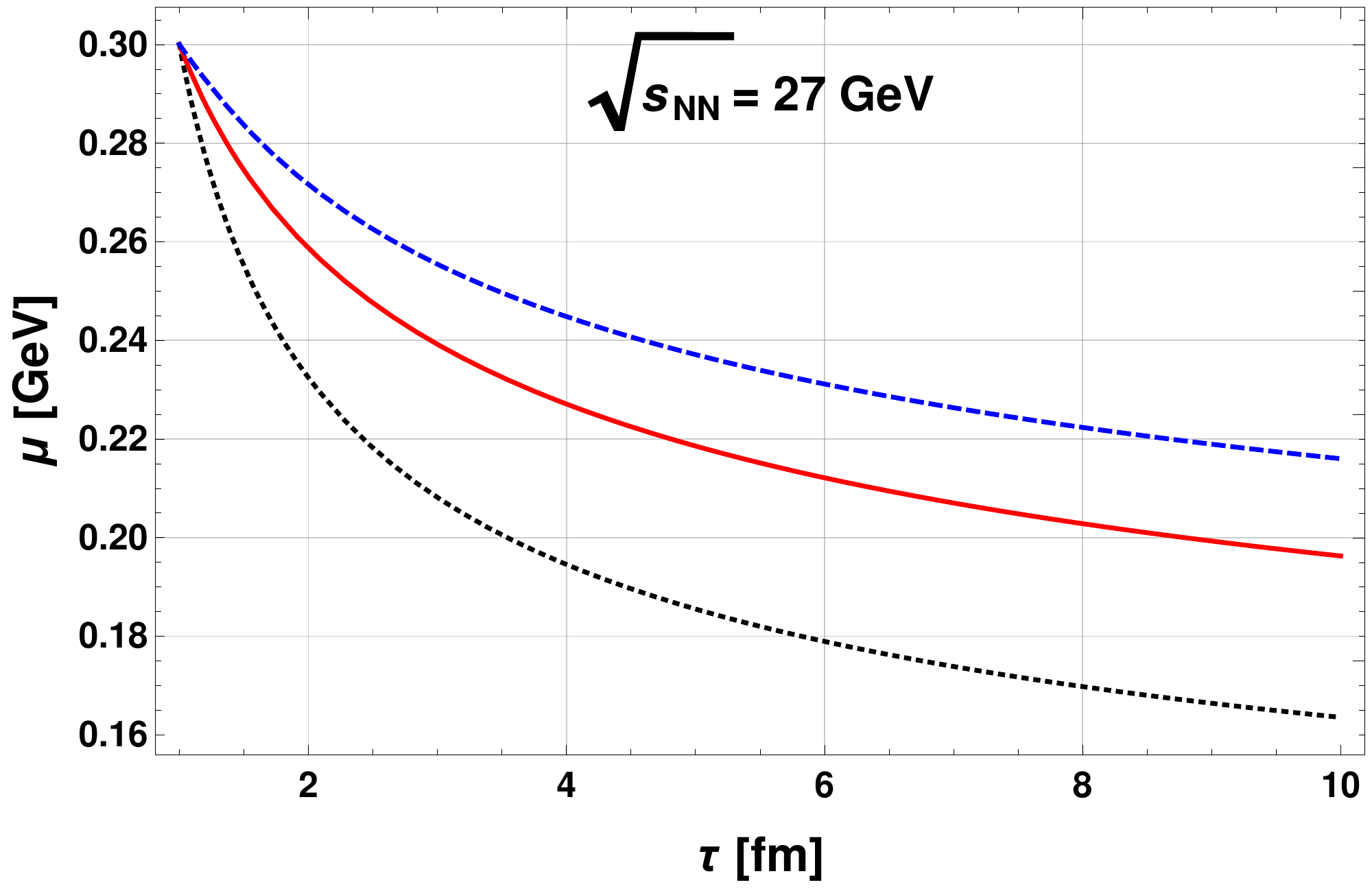}
\includegraphics[width=8.5cm]{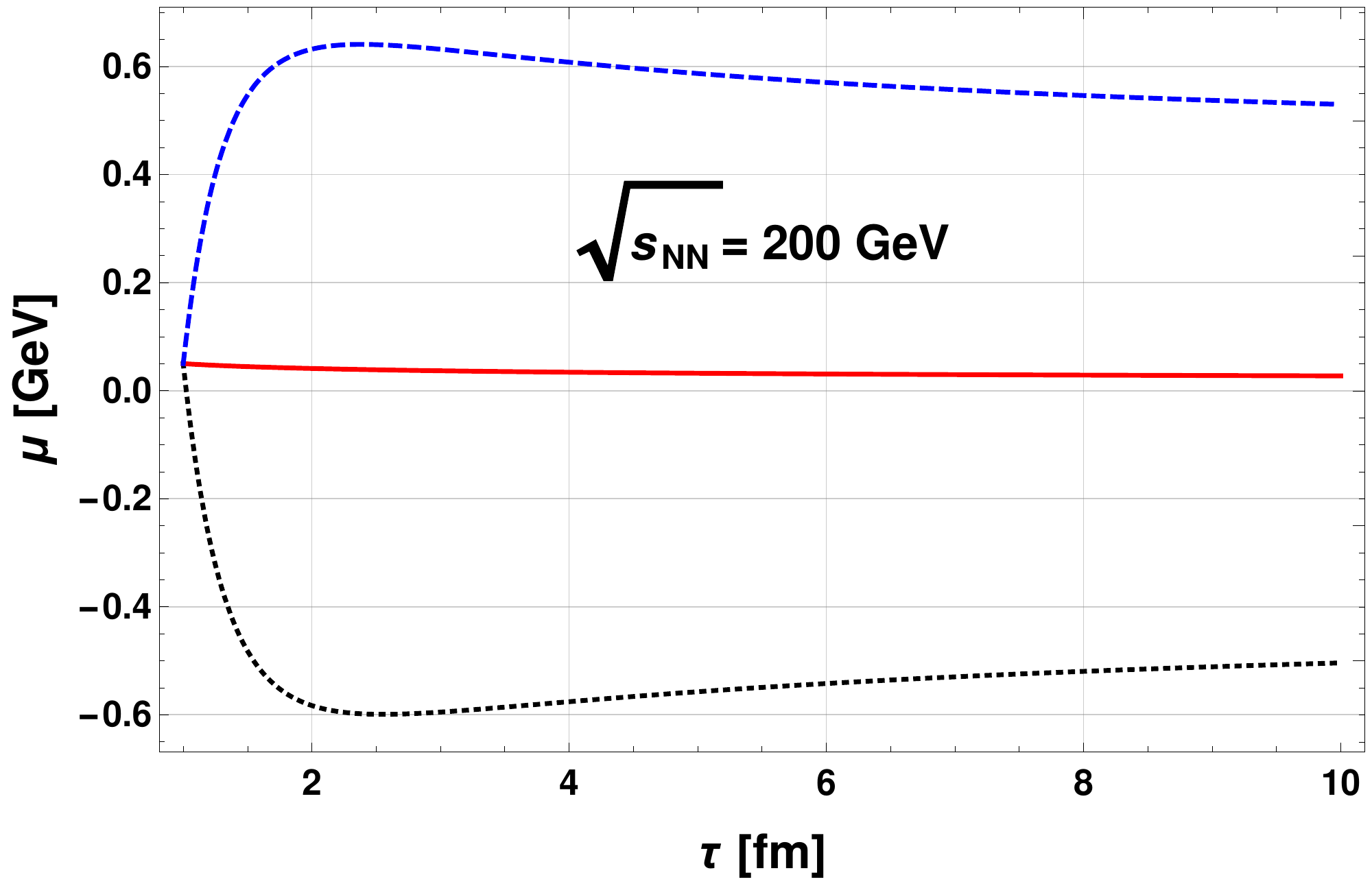}
\end{center}
\caption{\small Baryon chemical potential evolution with initial baryon chemical potential $\mu_{0} = 300\MeV$ for $\sqrt{s_{\rm NN}}=27\,{\rm GeV}$ (upper panel) and $\mu_{0} = 50\MeV$ for $\sqrt{s_{\rm NN}}=200\,{\rm GeV}$ (lower panel).}
\label{fig:Mu}
\end{figure}
\begin{figure}[t]
\begin{center}
\includegraphics[width=8.5cm]{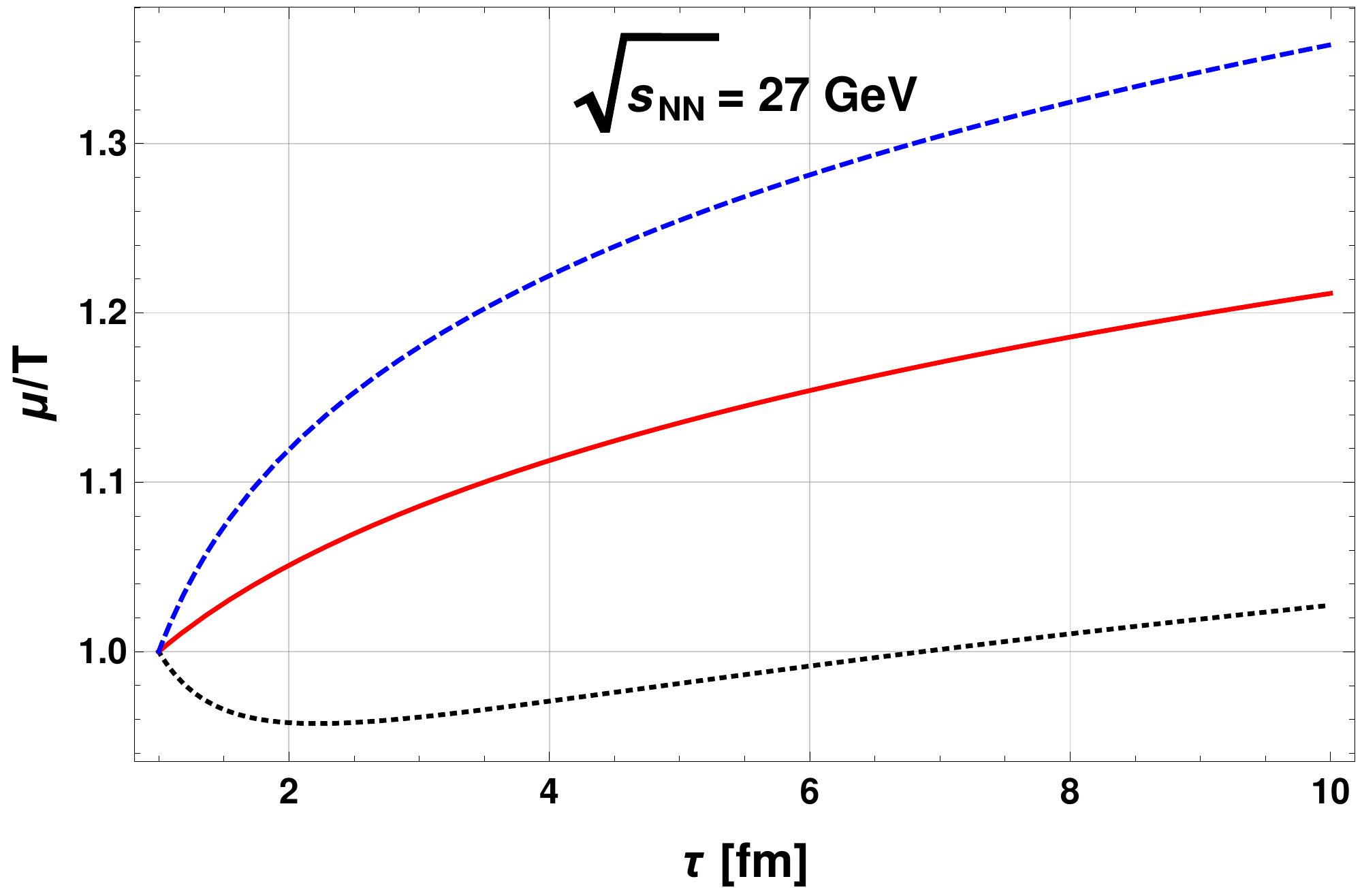}
\includegraphics[width=8.5cm]{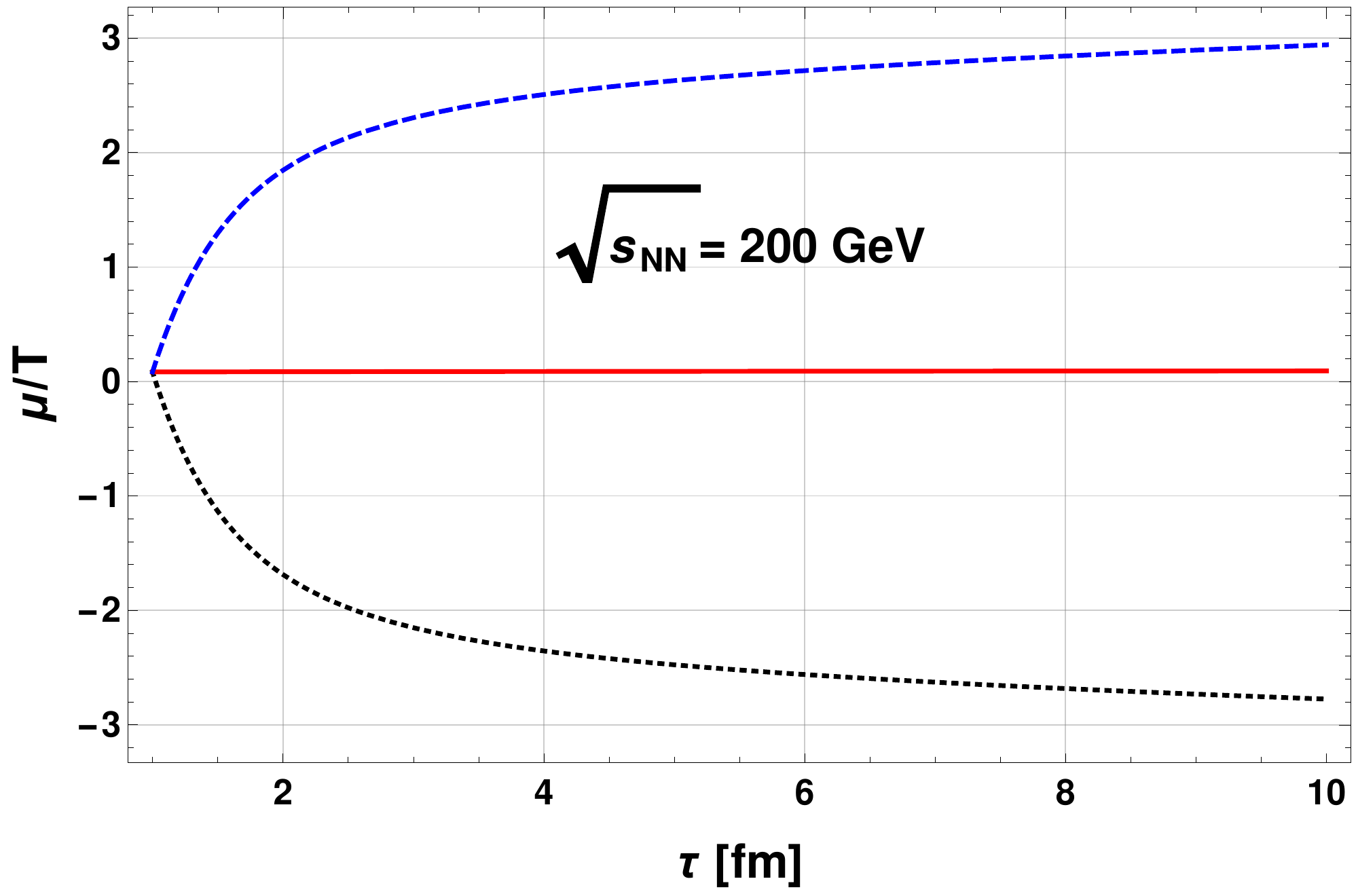}
\end{center}
\caption{\small Ratio of baryon chemical potential over temperature for  $\sqrt{s_{\rm NN}}=27\,{\rm GeV}$ (upper panel) and $\sqrt{s_{\rm NN}}=200\,{\rm GeV}$ (lower panel).}
\label{fig:MT}
\end{figure}
\begin{figure}[t]
\begin{center}
\includegraphics[width=8.5cm]{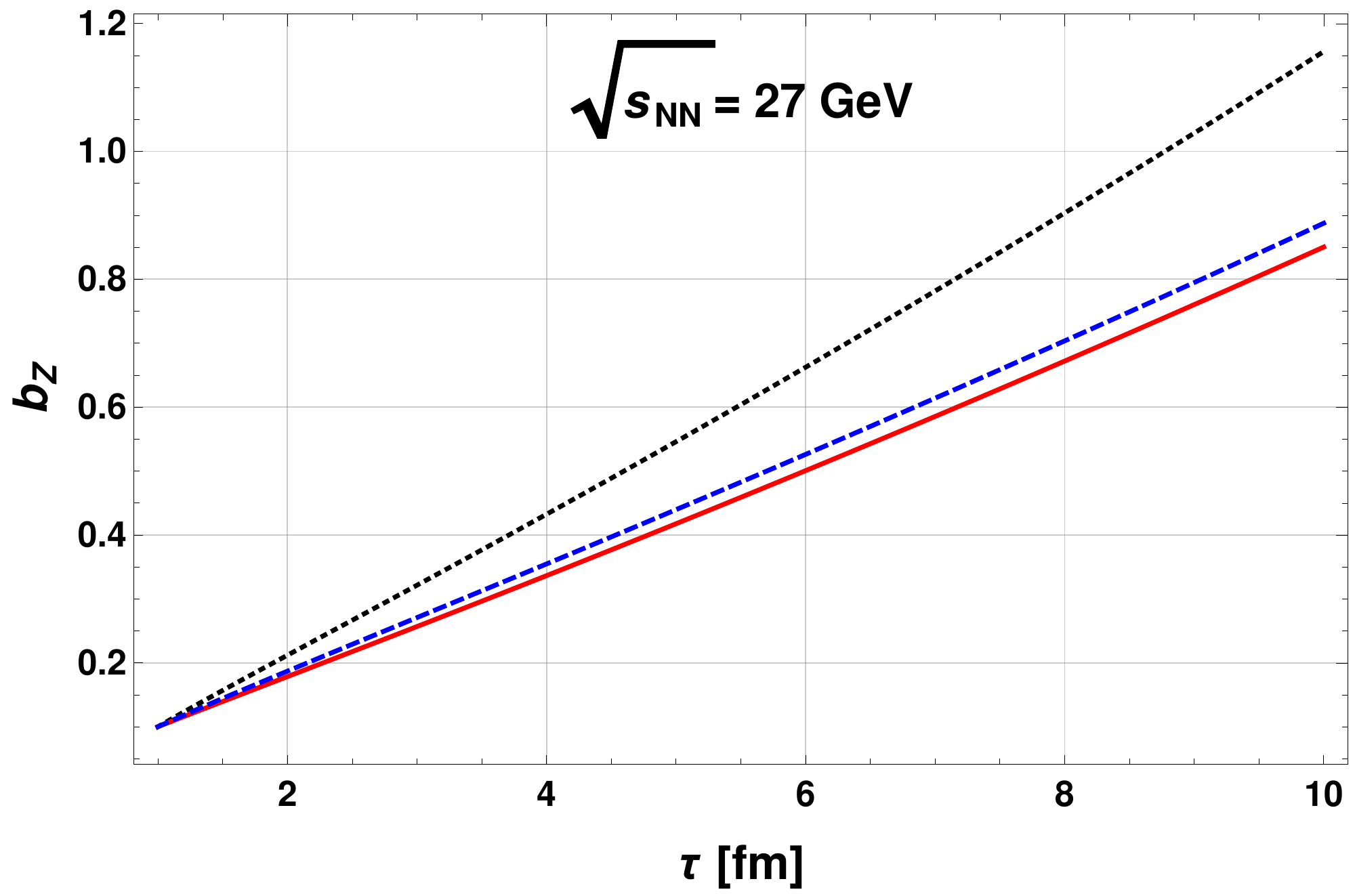}
\includegraphics[width=8.5cm]{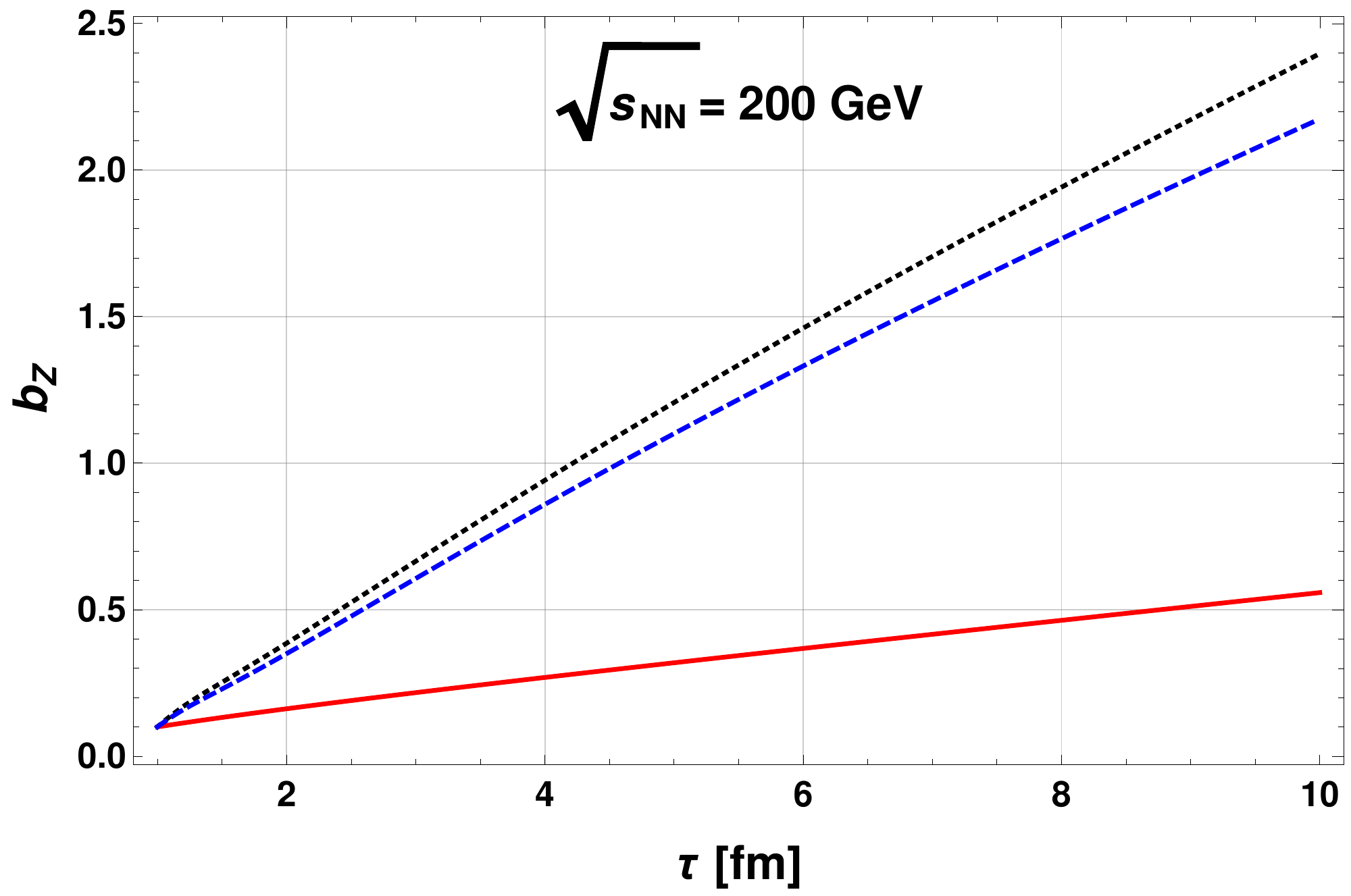}
\end{center}
\caption{\small Spin polarization coefficient $b_Z$ profile for  $\sqrt{s_{\rm NN}}=27\,{\rm GeV}$ (upper panel) and $\sqrt{s_{\rm NN}}=200\,{\rm GeV}$ (lower panel) with initial value $b_{Z}^0=0.1$. The modification of the $b_{Z}$ evolution slope due to electric field is much more pronounced when $\mu_0/T_0$ is small as can be seen in the lower panel.}
\label{fig:Bz}
\end{figure}

As the above form suggests, the evolution of energy density is determined by the competition between the expansion term, $\sim w$, and Joule heating (JH) term. The electric field plays two opposite roles. Firstly, it produces entropy and therefore increases the temperature compared to the case without an electric field. To realize this, assume an uncharged conformal fluid with the equation of state (EOS), $\varepsilon=3P=3f_PT^4$, with $f_P$ being a pure number~\cite{Florkowski:2017olj}. Then \rf{eq:competition} is transformed to
\be
\frac{{d\log T}}{{d\log\tau}} = -\frac{1}{3}+\frac{\cond\tau}{12f_P} \Bigg(\frac{E}{T^2}\Bigg)^2.
\ee
The second term on the right hand side increases the temperature and the heating effect gets enhanced by increasing the values of $E/T^2$ and $\cond\tau$. However, according to \rf{eq:conductivity}, the factor $\cond\tau$ is a small number during the hydrodynamic evolution, and increasing it suppresses the electric field, see \rf{eq:em-fields}.
On the other hand, if the initial electric field is sufficiently large, the JH term may still dominate over expansion in early times. In such a case, a reheating effect is possible \cite{Shokri:2017xxn}. Nevertheless, the electric field in our setup modifies the dynamics of fluid not only through the JH term but also through the EOS. Consequently, the analytical investigation presented in \rfc{Shokri:2017xxn} for the reheating conditions is not applicable. Still, some important observations can be made using numerical inspections, such as studying the $\tau$-derivatives of $T$ and $\mu$ at the initial time for various values of $\alpha$, see \rff{fig:derivatives}, which shows that the reheating observed in \rfcs{Roy:2015kma,Shokri:2017xxn} is not possible in our setup.
For an early-time reheating to occur at a particular value of $\alpha\neq 0$, the initial derivative of $T$ must be positive. It means that if the initial derivative of $T$ is larger than the derivative at $\alpha=0$, then the JH effect comes into play making the fluid hotter. Inversely, if the initial derivative of $T$ is smaller than the derivative at $\alpha=0$, then the electric field is making the fluid cooler.
This \emph{electric cooling effect} occurs because the electric field makes the fluid elements heavier which in consequence increases their enthalpy density, as can be seen in \rff{fig:hplot}. For small $\mu/T$, this effect can be seen using a Taylor expansion of $w$ in $\mu/T$ which reads
\ba
w_i&=&4\cosh(q_i\ell \frac{ER_{\rm RMS}}{T})\Bigg(1-\tanh(q_i\ell \frac{ER_{\rm RMS}}{T})\frac{\mu}{3T}\nn\\
&& \hspace{1cm}+\half\left(\frac{\mu}{3T}\right)^2+\order{\frac{\mu}{3T}}^3\Bigg)\,w_{(0),i}(T)\,,
\label{eq:small-mu-exp}
\ea
where $w_{(0),i}=\varepsilon_{(0),i}+P_{(0),i}$. Here, the overall factor $\cosh(q_i\ell \frac{ER_{\rm RMS}}{T})$ is larger than one. The second term in the parenthesis is a quantity with an absolute value smaller than one with the sign given by ${\rm sgn}(q_i\ell)$. Therefore, for $q_i\ell<0$ the electric field enhances the enthalpy density slightly more than in the opposite case, see \rff{fig:hplot}. Since there are more negatively charged quarks than positively charged ones in $N_f=3$ case, for $\ell<0$, the enthalpy density is larger which can be seen in lower collision energies where $\mu/T$ is of order unity. However, this difference is negligible in small $\mu/T$ regime. Consequently, the temperature dynamics is similar for both negative and positive values of $\alpha$, as can be seen in the lower panel of \rff{fig:T}. This larger enthalpy density may result a lower temperature for $\alpha<0$. Nevertheless, this is not pronounced unless $\alpha$ is extremely large. For moderate values of $\alpha$ shown in the plots, the asymmetry of \rff{fig:hplot} is not rendered into asymmetry of temperatures for positive and negative $\alpha$, as is seen in the upper panel of \rff{fig:T}.

By further inspection of the initial derivative of $T$, we can understand the relation between the value of $\alpha$ and electric cooling effect. As it turns out, there is an interval around $\abs{\alpha}=0$, for which increasing $\abs{\alpha}$ enhances the electric cooling effect (see the  peak around $\alpha=0$ in the upper panel of \rff{fig:derivatives}). When $\abs{\alpha}$ reaches a threshold, i.e., the two minima in the upper panel of \rff{fig:derivatives}, this behavior is turned opposite, and the temperature starts rising with $\abs{\alpha}$. This is because the JH term is getting strong enough to partially counterbalance the electric cooling effect. As \rff{fig:T} suggests, the electric field effects in the dynamics of temperature is less pronounced for larger values of $\mu/T$, since $\mu$ dominates over $E$. We should emphasize that, from a phenomenological perspective, the occurrence of early-time reheating is unlikely. If existed, such a reheating should have been already observed, for instance, via electromagnetic probes.

As the lower panel of \rff{fig:derivatives} indicates, the dynamics of chemical potential is much more sensitive to the electric field than the dynamics of temperature. This is confirmed by \rff{fig:Mu}.
In the small $\mu/T$ regime (the lower panel in \rff{fig:Mu}), the late-time absolute value of $\mu$ is always larger for $\alpha \neq 0$ case, and $\mu$ has the same sign as $\alpha$. On the other hand, for $\mu/T$ of order unity (the upper panel in \rff{fig:Mu}) $\mu$ dominates over the electric field and the change of sign requires very large electric fields. Consequently, if $\alpha$ is not too large, $\mu$ is enhanced (suppressed) for positive (negative) values of $\alpha$. Moreover, as the results on $\mu/T$ suggest, see Fig.~(\ref{fig:MT}), the electric field modifies the trajectory which the fireball passes through the QCD phase diagram.

Finally, we find that the behavior of the spin polarization components, see Fig.~(\ref{fig:Bz}), is qualitatively similar to the case of pure Bjorken-expanding perfect-fluid background without electric field~\cite{Florkowski:2019qdp}, however, introducing the external electric field in the perfect-fluid background solely, interestingly enhances the dynamics of spin polarization coefficients. We also observe that the enhancement of the spin polarization coefficients depends on the ratio $\mu_0/T_0$. Slope rises drastically for the small value of $\mu_0/T_0$, i.e., in the higher beam energy case. Hence, our results suggest that the external electric field in the background may play an important role in the polarization dynamics in heavy-ion collisions.
In this work we present only the behavior of $b_Z$ component, however all other spin polarization components exhibit similar qualitative features.

The spin polarization coefficients studied in this work is directly related to the  spin polarization of the hyperon's emitted from the system at the freeze-out through the formula for the mean spin polarization per particle $\langle\pi_{\mu}(p)\rangle$~\cite{Florkowski:2018ahw,Florkowski:2019qdp}
\beq
\langle\pi_{\mu}\rangle=\frac{E_p\frac{d\Pi _{\mu }(p)}{d^3 p}}{E_p\frac{d{\cal{N}}(p)}{d^3 p}}.
\label{mean}
\eeq
The above equation is the ratio of the invariant momentum distribution of the total Pauli-Luba\'nski vector and the momentum density of particles and antiparticles expressed as
\begin{equation}
E_p\frac{d\Pi _{\mu }(p)}{d^3 p} = \f{ \cosh(\bfug)}{(2 \pi )^3 m}
\int
\Delta \Sigma _{\lambda } p^{\lambda } \,
e^{-\beta \cdot p} \,
\tilde{\omega }_{\beta \mu }~p^{\beta }, \label{PDPLV}
\end{equation}
and
\beq
E_p\frac{d{\cal{N}}(p)}{d^3 p}&=&
\f{4 \cosh(\bfug)}{(2 \pi )^3}
\int
\Delta \Sigma _{\lambda } p^{\lambda } 
\,
e^{-\beta \cdot p} \,,
\eeq
 respectively, where $\tilde{\omega }^{\mu \nu }= (1/2) \epsilon^{\mu\nu\alpha\beta}  {\omega}_{\alpha\beta}$ is the dual polarization tensor~\cite{Florkowski:2019qdp} and $\Delta \Sigma _{\lambda }$ is the infinitesimal element of the freeze-out hypersurface.

We find that due to the enhancement of the spin polarization coefficients, the dynamics of the average spin polarization of $\Lambda$ hyperons, Eq. \eqref{mean}, also gets enhanced and exhibit similar qualitative behavior as compared to the case without electric field in the perfect-fluid Bjorken-expanding background studied in Ref.~\cite{Florkowski:2019qdp}.
\section{Summary and conclusions}
\label{sec:summary}
In the present work, we have studied the spin polarization dynamics in a Bjorken-expanding perfect-fluid resistive MHD background. In equilibrium, we used the stationary solution to the Boltzmann-Vlasov equation to find the modification of the baryon density, energy density, pressure, and energy-momentum tensor in the presence of an external electric field. We have assumed that the fluid is described by the BV solution for a medium composed of non-interacting quark-like quasi-particles. The latter results were used in the MHD equations. From the Maxwell equations and the Bjorken symmetry the evolution of \EM fields is readily determined. The symmetry implies that the electric and magnetic field in the LRF must be either parallel or antiparallel and oriented in the transverse directions. In such a solution, there is no extra source of fluid acceleration due to the Poynting term. 

The MHD equations in our case are reduced to the energy-momentum conservation and the baryon number conservation. We have solved these two equations numerically to find the evolution of temperature and baryon chemical potential and found that the electric field plays a twofold role in this regard. On the one hand, the electric field produces entropy through the Joule heating (JH) term and therefore increases the temperature. On the other hand, it makes the fluid enthalpy density larger, which results in a faster decrease of the temperature. For the collision energies that we have studied, this electric cooling effect is always dominant over the JH term. Therefore our results show a faster decay of temperature than in the purely hydrodynamic case. The dynamics of baryon chemical potential depends on the initial value of $\mu/T$. For small values of $\mu_0/T_0$, the parallel and anti-parallel \EM field cases are almost symmetric, and the baryon chemical potential changes sign with the sign of $\alpha$, while its absolute value is always larger than in the purely hydrodynamic case ($\alpha=0$). On the other hand, for larger values of $\mu_0/T_0$, the evolution of baryon chemical potential is different for parallel and anti-parallel configurations. The anti-parallel field cases (i.e., for $\alpha<0$) have lower temperature and the baryon chemical potential decreases faster than the purely hydrodynamic case ($\alpha=0$), while the baryon chemical potential of the parallel fields (i.e., for $\alpha>0$) are larger than in the purely hydrodynamic case. 

The dynamics of temperature and baryon chemical potential is used to solve the spin conservation law. The resulting spin polarization dynamics has qualitatively similar behavior as in the case of perfect-fluid  Bjorken-expanding background without electric field, but we observe that the dynamics of the spin polarization coefficients is enhanced due to the presence of electric fields in the background. It was found that for small values of $\mu_0/T_0$, the slope is steeper, and behavior gets enhanced significantly. Therefore, it suggests that the electric field may play a significant role in the polarization dynamics of $\Lambda$ hyperons if it is sufficiently large.

In the present work, the simplest available solution to the resistive MHD was employed, in which the symmetries of Bjorken flow are fully preserved. This implies that the qualitative behavior of the spin polarization remains unchanged. In a more realistic setup, one may consider the modification of the flow and breakdown of the symmetries induced by the \EM fields, which may change the dynamics of the spin polarization --  we leave these problems for future investigations. Other possible extension of the present work is to include the coupling between the \EM fields and spin degrees of freedom at the microscopic level. Investigations along these lines are ongoing and will be reported elsewhere.
\begin{acknowledgments}
We thank W. Florkowski, I. Karpenko, D. Rischke, N. Sadooghi and A. Tabatabaee for fruitful discussions. This research was supported in part by the Polish National Science Centre Grants No. 2016/23/B/ST2/00717 and No. 2018/30/E/ST2/00432. 
\end{acknowledgments}
\bibliography{pv_ref}{}
\bibliographystyle{utphys}
\end{document}